\documentstyle[epsfig]{elsart}
\begin{document}
\begin{frontmatter}
\title{A dynamical model describing stock market price distributions}
\author[ub]{Jaume Masoliver},
\author[ub]{Miquel Montero} and
\author[ub,gaesco]{Josep M. Porr\`a}.
 \address[ub]{Departament de F\'{\i}sica Fonamental, 
Universitat de Barcelona, Diagonal, 647, 08028-Barcelona, Spain}
\address[gaesco]{Gaesco Bolsa, SVB, S.A., Diagonal, 429, 08036-Barcelona,
Spain} 
\date{\today}
\maketitle

\begin{abstract}
High frequency data in finance have led to a deeper understanding on
probability distributions of market prices. Several facts seem to be well
stablished by empirical evidence. Specifically, probability distributions
have the following properties: (i) They are not Gaussian and their center
is well adjusted by L\'evy
distributions. (ii) They are long-tailed but
have finite moments of any order. (iii) They are self-similar on many time
scales. Finally, (iv) at small time scales, price volatility follows a
non-diffusive behavior.  We extend Merton's ideas on speculative price
formation and present a dynamical model resulting in a
characteristic function that explains in a natural way all of the above
features. The knowledge of such distribution opens a new and useful way of
quantifying financial risk. The results of the model agree ---with high
degree of accuracy--- with empirical data taken from historical records of
the Standard \& Poor's 500 cash index.
\end{abstract}

\end{frontmatter}

\section{Introduction}
One of the most important problems in mathematical finance is to
know the probability distribution of speculative prices. In spite
of its importance for both theoretical and practical applications the
problem is yet unsolved. The first approach to the problem
was given by Bachelier in 1900 when he modelled price dynamics as an
ordinary random walk where prices can go up and down due to a variety of
many independent random causes. Consequently the
distribution of prices was Gaussian~\cite{cootner}. The normal
distribution is ubiquitous in all branches of natural
and social sciences and this is basically due to the Central Limit
Theorem:
the sum of independent, or weakly dependent, random disturbances, all of
them with finite variance, results in a Gaussian random variable. Gaussian
models are thus widely used in finance although, as Kendall first
noticed~\cite{kendall}, the normal
distribution does not fit financial data specially at the wings of the
distribution. Thus, for instance, the
probability of events corresponding to 5 or more standard deviations is
around $10^4$ times larger than the one predicted by the Gaussian
distribution, in other words, the empirical distributions of prices are
highly leptokurtic. Is the existence of too many of such events, the so
called outliers, the reason for the existence of ``fat tails" and the
uselessness of the normal density specially at the wings of the
distribution. Needless to say that the tails of the price distributions
are crucial in the analysis of financial risk. Therefore, obtaining a
reliable distribution has deep consequences from a practical point of
view~\cite{lo,bouchaud}. 

One of the first attempts to explain the appearance of long tails in
financial data was taken by Mandelbrot in 1963~\cite{mandelbrot} who,
based on Pareto-L\'evy stable laws~\cite{feller}, obtained a leptokurtic
distribution. Nevertheless, the price to pay is high: the
resulting probability density function has no finite moments, except 
the first one. This is
indeed a severe limitation and it is not surprising since Mandelbrot's
approach can still be considered within the framework of the Central Limit
Theorem, that is, the sum of independent random disturbances of infinite
variance results in the L\'evy distribution which has infinite
variance~\cite{feller}. On the other hand, the L\'evy distribution has
been
tested against data in a great variety of situations, always with the same
result: the tails of the distribution are far too long compared with
actual data. In any case, as Mantegna and Stanley have recently
shown~\cite{mantegna1}, the L\'evy distribution fits very well
the center of empirical distributions ---much better than the Gaussian
density--- and it also shares the scaling behavior shown in
data~\cite{mantegna1,scalas,gash,gallu}. 

Therefore, if we want to explain speculative price dynamics as a sum of
weakly interdependent random disturbances, we are confronted with two
different and in some way opposed situations. If we assume finite
variance the tails are ``too thin" and the resulting Gaussian distribution
only accounts for a narrow neighborhood at the center of the
distribution. On the other hand, the assumption of infinite variance leads
to the L\'evy distribution which explains quite well a wider neighborhood
at the center of distributions but results in ``too fat tails". The
necessity of having an intermediate model is thus clear and this is the
main objective of the paper. 

Obviously, since the works of Mandelbrot~\cite{mandelbrot} and
Fama~\cite{fama} on L\'evy distributions, there have been several
approaches
to the problem, some of them applying cut-off procedures of the L\'evy
distribution~\cite{mantegna4,koponen} and, more recently, the use of ARCH
and GARCH models to obtain leptokurtic distributions~\cite{bolle92}.
The approaches based on cut-off procedures are approximations to the
distributions trying to better fit the existing data, but they are not
based on a dynamical model that can predict their precise features. On the
other hand ARCH~\cite{engle} and GARCH~\cite{bollerslev} models are indeed
dynamical adaptive models but they do not provide an overall picture of
the market dynamics resulting in a distinctive probability distribution.
In fact, ARCH/GARCH models usually assume that the market is Gaussian with
an unknown time-varying variance so to be self-adjusted to obtain
predictions. 

The paper is organized as follows. In Sect. 2 we propose the stochastic
model and set the mathematical framework that leads to a probability
distribution of prices. In Sect. 3 we present the main results achieved by
the model. Conclusions are drawn in Sect. 4.

\section{Analysis}

Let $S(t)$ be a random processes representing stock prices or some market
index value. The usual hypothesis is to assume that $S(t)$ obeys an
stochastic differential equation of the form 
\begin{equation}
\dot{S}/S=\rho+F(t),
\label{0}
\end{equation}
where $\rho$ is the instantaneous expected rate of return and $F(t)$ is
a random process with specified statistics, usually $F(t)$ is
zero-mean Gaussian white noise, $F(t)=\xi(t)$, in other words
$dW(t)=\xi(t)dt$, where $W(t)$ is the Wiener process or Brownian motion.
In this case, the dynamics of the market is clear since the return
$R(t)\equiv\log[S(t)/S(0)]$ obeys the equation 
$\dot{R}=\rho+\xi(t)$ which means that returns evolve like an overdamped
Brownian particle driven by the ``inflation rate" $\rho$ and, in
consequence, the return distribution is Gaussian. 

Let us take a closer look at the price formation and dynamics.
Following Merton~\cite{merton} we say that the change in the stock price
(or index) is
basically due to the random arrival of new information.
This mechanism is assumed to produce a marginal change in the price and it
is modelled by the standard geometric Brownian motion defined above. In
addition to this ``normal vibration" in price, there is an ``abnormal
vibration" basically due to the (random) arrival of important new
information that has more than a marginal effect on price. Merton models
this mechanism as a jump process with two sources of randomness: the
arrival times when jumps occurs, and the jump amplitudes. The result on
the overall picture is that the noise source $F(t)$ in price equation is
now formed by the sum of two independent random components
\begin{equation}
F(t)=\xi(t)+f(t),
\label{noise1}
\end{equation}
where $\xi(t)$ is Gaussian white noise corresponding
to the normal vibration, and $f(t)$ is ``shot noise" corresponding to the
abnormal vibration in price. This shot noise component can be explicitly
written as 
\begin{equation}
f(t)=\sum_{k=1}^\infty A_k\delta(t-t_k),
\label{shot}
\end{equation}
where $\delta(t)$ is the Dirac delta function, $A_k$ are jump amplitudes,
and $t_k$ are jump arrival times. It is also assumed that $A_k$ and $t_k$
are independent random variables with known probability distributions
given by $h(x)$ and $\psi(t)$ respectively~\cite{masoliver87}.

We now go beyond this description and specify the ``inner components" of
the normal vibration in price, by unifying this with Merton's abnormal
component. We thus assume that {\it all changes in the stock price (or
index) are modelled by different shot-noise sources corresponding to the
detailed arrival of information}, that is, we replace the total noise
$F(t)$ by the sum 
\begin{equation}
F(t)=\sum_{n=n_0}^{m}f_n(t),
\label{0a}
\end{equation}
where $f_n(t)$ are a set of independent shot-noise processes given by 
\begin{equation}
f_n(t)=\sum_{k_n=1}^\infty A_{k_n,n}\delta(t-t_{k_n,n}).
\label{0b}
\end{equation}
The amplitudes $A_{k_n,n}$ are independent random
variables with zero mean and probability density function (pdf),
$h_n(x)$, depending only on a single ``dimensional" parameter which,
without loss of generality, we assume to be the standard deviation of
jumps $\sigma_n  $, {\it i.e.}, 
\begin{equation}
h_n(x)=\sigma_n^{-1}h(x\sigma_n^{-1}).
\label{amplitude}
\end{equation}
We also assume that the occurrence of jumps is a Poisson process, in this
case shot noises are Markovian, and the pdf for the time interval between
jumps is exponential:
\begin{equation}
\psi(t_{k_n,n}-t_{k_n-1,n})=
\lambda_n\exp[-\lambda_n(t_{k_n,n}-t_{k_n-1,n})],
\label{psi}
\end{equation}
where $\lambda_n$ are mean jump frequencies, {\it i.e.}, $1/\lambda_n$ is
the
mean time between two consecutive jumps~\cite{masoliver87}. Finally, we
order the mean frequencies in a decreasing way: $\lambda_n<\lambda_{n-1}$.  

Let $X(t)$ be the zero-mean return, {\it i.e.}, $X(t)\equiv R(t)-\rho t$.
For our model $X(t)$ reads
\begin{equation}
X(t)=\sum_{n=n_0}^m\sum_{k_n=1}^\infty A_{k_n,n}\theta(t-t_{k_n,n}),
\label{6}
\end{equation}
where $\theta(t)$ is the Heaviside step function. Our
main objective is to obtain an expression for the pdf of
$X(t)$, $p(x,t)$, or equivalently, the characteristic
function (cf) of $X(t)$, $\tilde{p}(\omega,t)$, which is the Fourier
transform of the pdf $p(x,t)$. Note that $X(t)$ is a sum of independent
jump processes, this allows us to generalize Rice's method for a single
Markov shot noise to the present case of many shot noises~\cite{rice}. The
final result is
\begin{equation}
\tilde{p}(\omega,t)=\exp\left\{-t\sum_{n=n_0}^m
\lambda_n[1-\tilde{h}(\omega\sigma_n)]\right\}.
\label{7}
\end{equation}
As it is, $X(t)$ represents a shot noise process with mean frequency of
jumps given by $\lambda=\sum\lambda_n$ and jump distribution given by
$h(x)=\sum\lambda_nh_n(x)/\lambda$. Nevertheless, we make a
further approximation by assuming (i) $n_0=-\infty$, {\it i.e.}, there is
an infinite number of shot-noise sources, and (ii) there is no
characteristic time scale limiting the maximum feasible value of jump
frequencies, thus $\lambda_n\rightarrow\infty$ as $n\rightarrow-\infty$.
Both assumptions are
based on the fact that the ``normal vibration" in price is formed by 
the addition of (approximately) infinitely many random causes, which we
have modelled as shot noises. According to this, we introduce a
``coarse-grained" description and replace the sum in Eq.~(\ref{7}) by an
integral 
\begin{equation}
\tilde{p}(\omega,t)=\exp\left\{-t\int_{-\infty}^{u_m}
\lambda(u)[1-\tilde{h}(\omega\sigma(u))]du\right\}.
\label{7a}
\end{equation}
In order to proceed further we should specify a functional form for
$\lambda(u)$ and $\sigma(u)$. We note by empirical evidence
that the bigger a sudden market change is, the longer is the time we have
to wait until we observe it. Therefore, since $\lambda(u)$ decreases with
$u$ (recall that frequencies are decreasingly ordered) then $\sigma(u)$
must increase with $u$. We thus see that  $\sigma(u)$ has to be a positive
definite, regular and monotone increasing function for all $u$. The
simplest choice is: $\sigma(u)=\sigma_0e^{u}$. On the other hand, there is
empirical evidence of scaling properties in financial
data~\cite{mantegna1,scalas,gash,gallu}. We summarize the above
requirements ({\it i.e.}, inverse relation between $\lambda$ and $\sigma$,
and scaling) by imposing the ``dispersion relation": 
\begin{equation}
\lambda=\lambda_0(\sigma_0/\sigma)^{\alpha}.
\label{00}
\end{equation}
where $\alpha$ is the scaling parameter. Under these assumptions the cf of
the return $X(t)$ reads:
\begin{equation}
\tilde{p}(\omega,t)=\exp\left\{-\lambda_0t\sigma_0^\alpha
\int_{0}^{\sigma_m}z^{-1-\alpha}[1-\tilde{h}(\omega z)]dz\right\},
\label{11}
\end{equation}
where $\sigma_m=\sigma_0e^{u_m}$ is the maximum value of the standard
deviation. We observe that if $\sigma_m=\infty$, which means that some
shot-noise source has infinite variance, then Eq.~(\ref{11}) yields the
L\'evy distribution 
\begin{equation}
\tilde{L}_{\alpha}(\omega,t)=\exp(-kt\omega^\alpha),
\label{levy}
\end{equation}
where 
\begin{equation}
k=\lambda_0\sigma_0^\alpha \int_0^\infty
z^{-1-\alpha}[1-\tilde{h}(z)]dz.
\label{levyk}
\end{equation}
Hence, if we want a distribution with finite moments, we have to assume a
finite value for $\sigma_m$. 

Let $\lambda_m$ be the mean frequency corresponding
to the maximum (finite) variance. Recall that, in the discrete case ({\it
c.f.} Eq.~(\ref{7})), shot-noise sources are ordered, thus $\lambda_m$ and
$\sigma_m$ correspond to the mean frequency and the variance of the
last jump source considered. Our last assumption is that the total number
of noise sources in Eq.~(\ref{6}) increases with the observation time $t$
and, since $n_0=-\infty$, this implies that $m=m(t)$ is an increasing
function of time. Consequently, the mean period of the last jump source,
$\lambda_m^{-1}$, also grows with $t$. The
simplest choice is the linear relation: $\lambda_m t=a$, where $a>0$ is
constant. Therefore, from the dispersion relation, Eq.~(\ref{00}), we see
that the maximum jump variance depends on time as a power law: 
\begin{equation}
\sigma_m^2=(bt)^{2/\alpha},
\label{power}
\end{equation}
where $b\equiv\sigma_0^{\alpha}\lambda_0/a$. We finally have
\begin{equation}
\tilde{p}(\omega,t)=\exp\left\{-abt\int_{0}^{(bt)^{1/\alpha}}
z^{-1-\alpha}[1-\tilde{h}(\omega z)]dz\right\},
\label{final}
\end{equation}

\section{Results}

Let us now present the main results and consequences of the above
analysis. First,
the volatility of the return is given by
\begin{equation}
\langle X^2(t)\rangle=\frac{a\sigma_m^2}{2-\alpha}=
\frac{a}{2-\alpha}\ (bt)^{2/\alpha}, 
\label{13}
\end{equation}
which proves that $\alpha<2$ and the volatility shows 
super-diffusion. The anomalous diffusion behavior of the empirical data
(at least at small time scales) was first shown by Mantegna and Stanley
without mention it~\cite{mantegna3a,mantegna3}. Second, kurtosis is
constant and given by
\begin{equation}
\gamma_2=\frac{(2-\alpha)^2\tilde{h}^{(iv)}(0)}{(4-\alpha)a}.
\label{kurtosis}
\end{equation}
Thus $\gamma_2>0$ for all $t$, in other words, we have a leptokurtic
distribution in all time scales. Third, the return probability
distribution scales as
\begin{equation}
p(x,t)=(bt)^{-1/\alpha}p(x/(bt)^{1/\alpha})
\label{scale}
\end{equation} 
and the model becomes
self-similar~\cite{mantegna1,scalas,gash,gallu}.

\begin{figure}[htb]
\begin{center}
\epsfxsize=12 cm \epsffile{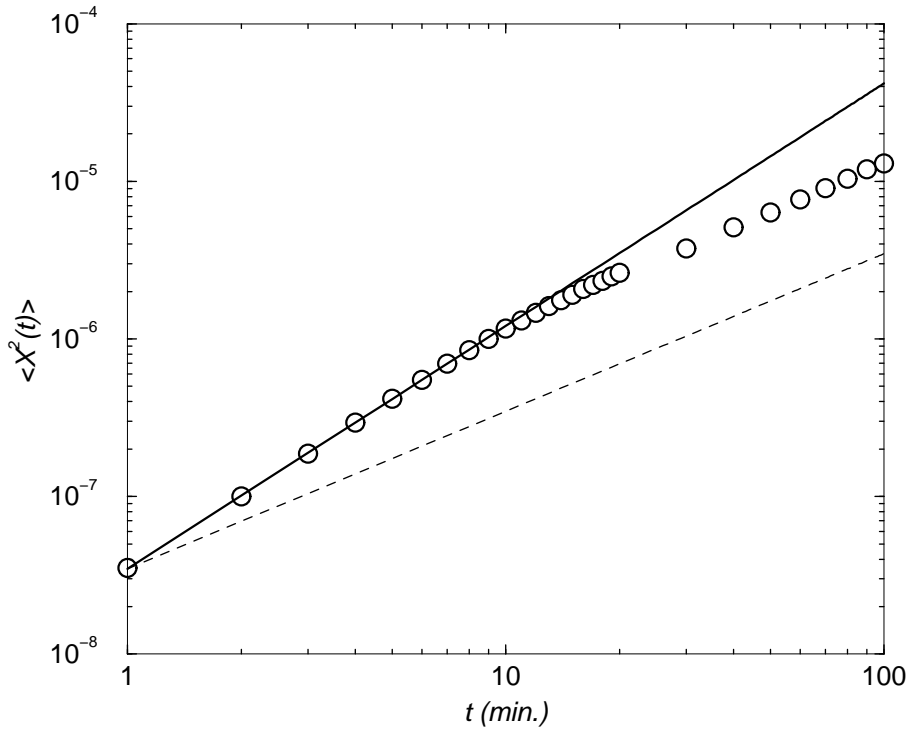}
\end{center}
\caption{Second moment of the zero-mean return. Circles
correspond to empirical data from S\&P 500 cash index (January 1988 to
December 1996). Solid line shows the super-diffusive character predicted
by Eq.~(\protect\ref{13}).}
\label{fig1}
\end{figure}

In Fig.~\ref{fig1} we plot the super-diffusion behavior. Circles
correspond to empirical data from S\&P 500 cash index during the period
January 1988 to December 1996. Solid line shows the super-diffusive
character predicted by Eq.~(\ref{13}) setting $\alpha=1.30$ and
$ab^{2/\alpha}=2.44\times 10^{-8}$ (if time is measured in minutes).
Dashed line represents normal-diffusion $\langle X^2(t)\rangle\propto t$.
Observe that data obeys super-diffusion for $t\leq 10$ min, and when
$t>10$ min there seems to be a ``crossover" to normal diffusion.

We finally study the asymptotic behavior of our distribution. It can be
shown from Eq.~(\ref{11}) that the center of the distribution, defined by
$|x|<(bt)^{1/\alpha}$, is again approximated by the L\'evy
distribution defined above. On the other hand the tails of the
distribution are solely determined by the jump pdf $h(u)$ by means of the
expression
\begin{equation}
p(x,t)\sim\frac{abt}{|x|^{1+\alpha}}
\int_{|x|/\sigma_m}^\infty u^\alpha
h(u)du,\qquad(|x|\gg (bt)^{1/\alpha}).
\label{14}
\end{equation}
Therefore, return distributions present fat tails and have finite moments
{\it if jump distributions behave in the same way}.  This, in turn, allows
us to make statistical hypothesis on the form of $h(u)$ based on the
empirical form and moments of the pdf. 

\begin{figure}[htb]
\begin{center}
\epsfxsize=12 cm \epsffile{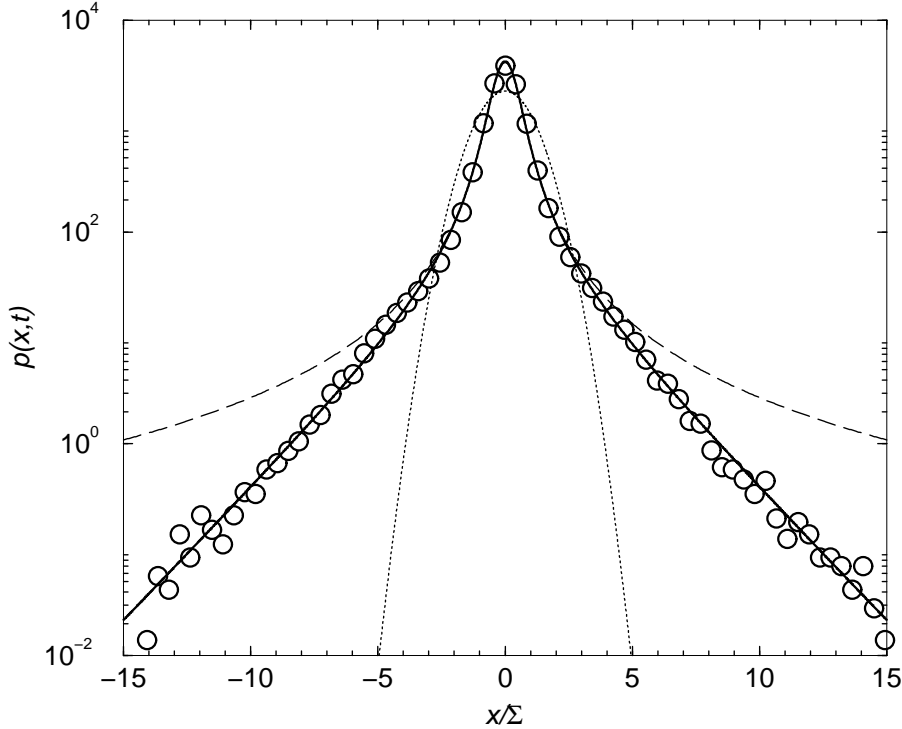}
\end{center}
\caption{Probability density function $p(x,t)$ for $t=1$ min. Circles
represent empirical data from S\&P 500 cash index (January 1988 to
December 1996). $\Sigma$ is the standard deviation of empirical data.
Dotted line corresponds to the Guassian density. Dashed line is the
L\'{e}vy distribution and the solid line is the Fourier inversion of
Eq.~(\protect\ref{last}) with a gamma distribution of jumps (see main text
for details).}
\label{fig2}
\end{figure}

In Fig.~\ref{fig2} we plot the
probability density $p(x,t)$ of the  S\&P 500 cash index returns
$X(t)$ observed at time $t=1$ min (circles).
$\Sigma=1.87\times 10^{-4}$ is the standard deviation of the empirical
data. Dotted line corresponds to a Gaussian density with standard
deviation given by $\Sigma$. Solid line shows the Fourier inversion of
Eq.~(\ref{11}) with $\alpha=1.30$, $\sigma_m=9.07\times 10^{-4}$, and
$a=2.97\times 10^{-3}$. We use the gamma distribution of the absolute
value of jump amplitudes,
\begin{equation}
h(u)=\mu^\beta|u|^{\beta-1}e^{-\mu|u|}/2\Gamma(\beta),
\label{beta}
\end{equation}
with $\beta=2.39$, and $\mu=\sqrt{\beta(\beta+1)}=2.85$. Dashed line
represents a symmetrical L\'evy stable distribution of index $\alpha=1.30$
and the scale factor $k=4.31\times 10^{-6}$ obtained from
Eq.~(\ref{levyk}).
We note that the values of $\sigma_m$ and $\Sigma$ predict that the
Pareto-L\'evy distribution fails to be an accurate description of the
empirical pdf for $x\gg 5\Sigma$ (see Eq.~(\ref{14})). 

We chose a gamma distribution of jumps because (i) as suggested by the
empirical data analized, the tails of $p(x,t)$ decay exponentially, and
(ii) one does not favor too small size jumps, {\it i.e.}, those jumps with
almost zero amplitudes. In any case, it would be very useful to get some
more microscopic approach (based, for instance, in a ``many agents"
model~\cite{lo,lux}) giving some inside on the particular form of $h(u)$.

\section{Conclusions}

Summarizing, by means of a continous description of random pulses,
we have obtained a dynamical model leading to a probability distribution
for the speculative price changes. This distribution which is given by the
following characteristic function:
\begin{equation}
\tilde{p}(\omega,t)=\exp\left\{-a\int_{0}^{1}
z^{-1-\alpha}[1-\tilde{h}(\omega z\sigma_m(t))]dz\right\},
\label{last}
\end{equation}
where $\sigma_m(t)=(bt)^{1/\alpha}$, it depends on three positive
constants: $a$, $b$, and $\alpha<2$. The characteristic
function~(\ref{last}) also depends on an unknown function
$\tilde{h}(\omega)$, the unit-variance characteristic function of jumps,
also to be conjectured and fitted from the tails of the empirical
distribution. Therefore, starting from simple and reasonable assumptions
we
have developed a new stochastic process that possesses many of the
features, {\it i. e.\/} fat tails, self-similarity, superdiffusion, and
finite moments,
of financial time series, thus providing us with a different point of view
on the dynamics of the market. We finally point out that the model does
not explain any correlation observed in empirical data (as some markets
seem to have~\cite{bouchaud,lo2}). This insufficiency is due to the fact
that we have modelled the behavior of returns through a mixture of
independent sources of white noise. The extension of the model to include
non-white noise sources and, hence, correlations will be presented soon.

\begin{ack}
This work has been supported in part by Direcci\'on General de
Investigaci\'on Cient\'{\i}fica y T\'ecnica under contract No. PB96-0188
and Project No. HB119-0104, and by Generalitat de Catalunya under contract
No. 1998 SGR-00015.
\end{ack}


\begin{thebibliography}{999}
\bibitem{cootner} {\it The Random Character of
Stock Market Prices}, P.H. Cootner ed. (MIT Press, Cambridge MA, 1964).
\bibitem{kendall} M.G. Kendall, J. Royal Stat. Soc. {\bf 96}, 11-25
(1953).
\bibitem{lo} J.Y. Campbell, A.W. Lo, and A.C. MacKinlay, {\it The
Econometrics of Financial Markets} (Princeton University Press, Princeton,
1997).
\bibitem{bouchaud} J.P. Bouchaud, and M. Potters,  {\it Th\'{e}orie des
Riskes Financiers} (Al\'{e}a-Saclay, Paris, 1997).
\bibitem{mandelbrot} B. Mandelbrot, J. Business {\bf 35}, 394-419 (1963).
\bibitem{feller} W. Feller, {\it An Introduction to Probability Theory and
its Applications} (J. Wiley, New York, 1971). 
\bibitem{mantegna1} R. N. Mantegna, and E.H. Stanley, Nature {\bf 376},
46-49 (1995).
\bibitem{scalas} E. Scalas, Physica A {\bf 253}, 394-402 (1998). 
\bibitem{gash} S. Gashghaie, W. Breymann, J. Peinke, P. Talkner, and
Y. Dodge, Nature {\bf 381}, 767-770 (1996).
\bibitem{gallu} S. Gallucio, G. Caldarelli, M. Marsili, and Y.-C. Zhang, 
Physica A {\bf 245}, 423-436 (1997).
\bibitem{fama} E. Fama, J. Business {\bf 35}, 420-429 (1963).
\bibitem{mantegna4} R.N. Mantegna, and E.H. Stanley, Phys. Rev. Lett. {\bf
73}, 2946-2949 (1994). 
\bibitem{koponen} I. Koponen, Phys. Rev. E {\bf 52}, 1197-1199 (1995).
\bibitem{bolle92} T. Bollerslev, R.Y. Chou, and K.F. Kroner, J.
Econometrics {\bf 52}, 5-59 (1992). 
\bibitem{engle} R.F. Engle, Econometrica {\bf 50}, 987-1007 (1982). 
\bibitem{bollerslev} T. Bollerslev, J. Econometrics {\bf 31}, 307-327
(1986). 
\bibitem{merton} R.C. Merton, J. Financial Economics {\bf 3}, 125-144
(1976).
\bibitem{masoliver87} J. Masoliver, Phys. Rev. A {\bf 35}, 3918-3928
(1987). 
\bibitem{rice} S.O. Rice, in {\it Noise and Stochastic Processes},  N. Wax
ed. (Dover, New York, 1954).
\bibitem{mantegna3a}  R.N. Mantegna, and E.H. Stanley, Nature {\bf 383},
587-588 (1996).
\bibitem{mantegna3} R. N. Mantegna, and E.H. Stanley, Physica
A {\bf 239}, 255-266 (1997).
\bibitem{lux} T. Lux, and M. Marchesi, Nature {\bf 397}, 498-500 (1999).
\bibitem{lo2} A. W. Lo, and A. C. MacKinlay, Rev. Financial Studies {\bf
1}, 41-66 (1988); {\bf 3}, 175-205 (1990).
\end{thebibliography}
\end{document}